\documentclass[preprint,12pt]{elsarticle}

\usepackage{amssymb}
\usepackage{multirow}
\usepackage{booktabs} % midrule and toprule
\usepackage{rotating}%sideways

\usepackage[hidelinks]{hyperref}
\usepackage{graphicx}
\usepackage{subcaption}

\usepackage{amsmath}
\journal{Arxiv (Information Systems Journal)}

\begin{document}

\begin{frontmatter}



\title{Quality-aware skill translation models for expert finding on StackOverflow}




%\author{Arash Dargahi Nobari}
%%\cortext[arash]{Corresponding author}
%\ead{ a.dargahinobari@mail.sbu.ac.ir}
%\address{Faculty of Computer Science and Engineering Shahid Beheshti University:G.C}
%
%\author{Mahmood Neshati}
%\ead{ m\_neshati@sbu.ac.ir}
%\address{Faculty of Computer Science and Engineering Shahid Beheshti University:G.C}

%\author[add1]{Arash Dargahi Nobari}
\author{Arash Dargahi Nobari}
\ead{a.dargahinobari@mail.sbu.ac.ir}
\author{Mahmood Neshati}
\ead{ m\_neshati@sbu.ac.ir}
\author{Sajad Sotudeh Gharebagh}
\ead{s.sotudeh@mail.sbu.ac.ir}

%\address[add1]{Faculty of Computer Science and Engineering Shahid Beheshti University:G.C}
\address{Faculty of Computer Science and Engineering, Shahid Beheshti University, G.C.}

\begin{abstract}
StackOverflow has become an emerging resource for talent recognition in recent years. While users exploit technical language on StackOverflow, recruiters try to find the relevant candidates for jobs using their own terminology. This procedure implies a gap which exists between recruiters and candidates terms. Due to this gap, the state-of-the-art expert finding models cannot effectively address the expert finding problem on StackOverflow. We propose two translation models to bridge this gap. The first approach is a statistical method and the second is based on word embedding approach. Utilizing several translations for a given query  during the scoring step, the result of each intermediate query is blended together to obtain the final ranking. Here, we propose a new approach which takes the quality of documents into account in scoring step. We have made several observations to visualize the effectiveness of the translation approaches and also the quality-aware scoring approach. Our experiments indicate the following: First, while statistical and word embedding translation approaches provide different translations for each query, both can considerably improve the recall. Besides, the quality-aware scoring approach can improve the precision remarkably. Finally, our best proposed method can improve the MAP measure up to 46\% on average, in comparison with the state-of-the-art expert finding approach.
\end{abstract}


\end{frontmatter}



\section{Introduction}
Nowadays community question answering (CQA) websites have gained a lot of interest among people due to their capabilities in solving different kinds of problems. Over the recent years, a swift growth in the number of users of these networks has been tracked. The popularity of these networks can be noticed by observing the traffic of the renowned CQA websites such as StackOverflow\footnote{stackverflow.com}, Quora\footnote{quora.com}, and Yahoo! Answers\footnote{answers.yahoo.com}. Currently, with the growing resource of information, CQA websites provide users a valuable platform for information sharing and searching \cite{tshaped}. Users can contribute and interact by posting questions and answers, commenting, voting, and etc. Additionally, in some CQAs such as StackOverflow they can mark the best answer among provided answers which bring about the concept of accepted answer.
\par
The vital key to the success of CQA platforms is the users who can provide high-quality answers to the more challenging questions posted in community \cite{derijke:2015:early}. In recent years, many studies have been made to address the expertise retrieval as a superior Information Retrieval (IR) task. Indeed, expert finding has recently attracted much attention in IR community \cite{Dargahi:2017:sigir,ZHOU2014136,Wei:2016:twitter} and become a well-studied field. The task of expert finding is defined as detecting a set of persons with relevant expertise for the given query \cite{deng:enhanced}.

Expert finding has many real-world applications. One of its trending applications is talent acquisition which benefits organizations significantly \cite{Neshati_ieee}. By analyzing historical data, recruiters can detect potential experts to evolve their organization's business. Moreover, as a part of revenue model, CQA platforms such as StackOverflow aim to find experts on different topics and then propose them to organizations \cite{sof_candidate,sof_job}. These examples demonstrate the high importance of expertise retrieval task.

As mentioned before, expert finding is a well-established study in the field of IR and simultaneously it is a challenging task. In recent years, several research studies have been conducted in different domains including CQA platforms \cite{Zhao:2015:ieee}, bibliographic networks \cite{Neshati_ieee}, and organizations \cite{karimzadehgan2009enhancing}. Evidently, the task of expert finding is not completely the same in these domains. Although some similarities exist between these domains, there are some remarkable disparities. The most critical affinity is that document associated with a candidate is the most noticeable evidence of his/her expertise on the subject of related topic and expertise level of the candidate can be estimated using some properties like vote (score) of the document. As in \cite{deng:enhanced}, the quality of a paper is estimated based on citation count of the paper. Nonetheless, in CQA platforms, the vote count of a document cannot merely show expertise of the author. Rather, it can also represent the popularity and novelty of the subject \cite{ravi2014great}. Moreover, in CQA platforms, the questions and the answers contain technical aspects of the language, hence in many cases, there is a deep gap between the main query and the language which is used by the expert people of the question-related community. As an illustration, consider the ``Android'' query. As expert users of the Android community do not use the term itself directly in their answers and instead they use some terms like ``fragment'' and ``broadcastreceiver'' which have a direct relation with ``Android'', we can say that a deep gap exists between the ``Android'' as main query and the language used by experts. 

In a typical CQA community, each question has one or more tags which indicate the required skills to answer that question. These tags can basically be considered as skill areas which recruiters are interested in. For example, consider the following question on StackOverflow, ``What is the difference between JPA and Hibernate?''. This question can be tagged by ``jpa'', ``hibernate'', ``java-ee'', and ``orm'' (i.e. Object Relational Mapping) which are important skill areas in Java programming language. In this paper, our goal is to detect and rank users who are skillful in a given skill area (tag) with respect to the title and body of the questions which are given.

The state-of-the-art models proposed by Balog \textit{et al.} \cite{balog:trends} can be used in order to rank expert users in StackOverflow. In these models, the question's associated tags can be considered as the main queries, as mentioned earlier, the body of the provided answers can be regarded as his/her evidence of author's expertise. The pitfall of these models is the vocabulary gap existed between the textual representation of skills (i.e. tags) and the body of answers provided by expert candidates. Indeed, these models fail to address vocabulary gap problem as they are based on exact and not semantic matching. Over the last few years, several models have been proposed to solve the vocabulary gap problem \cite{semantic}. Precisely, statistical translation models \cite{karimzadehgan:2010}, topic modeling approach \cite{Momtazi}, and more lately word embedding methods \cite{Dargahi:2017:sigir} which are among outstanding models to overcome this problem.

\par
In this paper, we propose two models to translate a given skill area (e.g. ``Android'', ``orm'' and etc.) to a set of relevant words. These translations can be useful to improve the matching between expert finding queries and the technical textual evidence (i.e. answers) associated with each candidate. They can also be used independently by recruiters to detect important aspects of each skill area. For example, “java-ee” skill area can be translated to \textit{application, web, spring, bean, service, http, session, request, controller} and \textit{ejb} which are important aspects of ``java-ee'' in StackOverflow.
Our first translation model (i.e. MI) is a statistical model based on mutual information and the second one (i.e. WE) is based on a word embedding method utilizing the specific structure and CQA's data to translate a skill area to its relevant words. After finding the appropriate translations, we have used four scoring approaches to combine the result of each translation to find the final ranking of experts for a given skill area.
\par
The basic idea of our work has been recently published as a short paper in SIGIR conference \cite{Dargahi:2017:sigir}. However, the conference paper does not include a complete description of the proposed algorithms due to the page limit. This paper is a significant extension of the published short paper. In this paper, we have made four significant contributions including 1) We have added a new dataset (i.e. ``PHP'' dataset) to evaluate our proposed models. 2) Deeper analysis of the results are conducted. 3) Having adopted translation models which bridge the vocabulary gap issue, we aimed to take the quality of documents into account and have improved results considerably using Voteshare based scoring approaches. 4) Our entire sources including code and datasets have been uploaded and are publicly available for researchers\footnote{http://tiny.cc/sofef}.
\begin{figure*}[!t]
	\centering
	\includegraphics[width=0.75\linewidth]{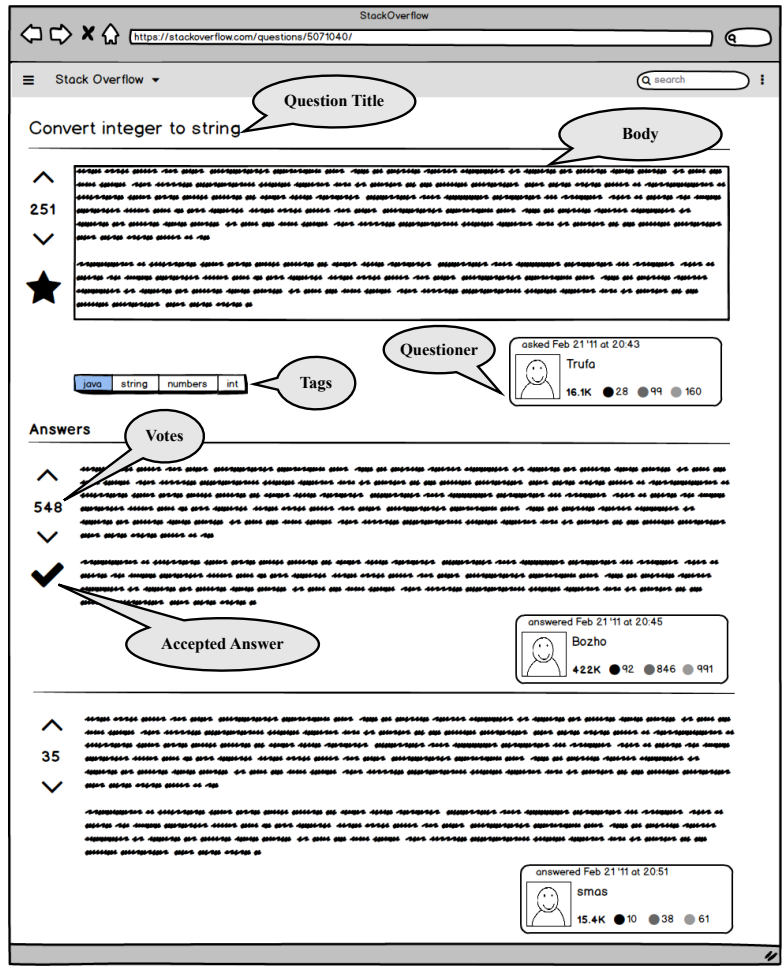}
	\caption{A sample question and its associated answers in StackOverflow. Title, body, tags, etc are highlighted. }
	\label{fig:sof}
\end{figure*}

\section{StackOverflow}
\label{sec:sof}
In this section, we briefly introduce the StackOverflow, its fundamentals and the major properties of its concepts including questions, answers, and user interactions.

As one of the most prominent CQAs, StackOverflow provides its users a quick access to expertise and knowledge. Users can ask questions, answer them, vote up or down and also edit posts. Some of these actions (e.g. voting up or down, commenting) are restricted to active members of the community. Users can gain or even lose reputation and badges with regard to their behavior and contribution to the community. For instance, the community of StackOverflow will reward users with 15 points, if their answers get accepted by the asker. Additionally, they can earn different levels of badges based on their high-quality contribution (e.g. an answer score of 100 or more will leads to ``Great Answer'' badge) which exemplifies gamification methods to motivate users.

Fig. \ref{fig:sof} demonstrates a sample question and the related answers on StackOverflow. Each post (i.e questions and answers) includes body, vote count (i.e. subtraction of down-votes from up-votes), a doer (i.e. questioner and answerer), a set of tags which have been chosen by the questioner, and possibly a set of comments by users. Besides, the view count of the post, the posting date and time As illustrated in Fig. \ref{fig:sof}. When a questioner feels satisfied with an answer, he/she can accept it. \textit{Acceptance} is determined by a green check-mark next to the accepted answer and is characterized as follows. \textit{``Accepting an answer is not meant to be a definitive and final statement indicating that the question has now been answered perfectly. It simply means that the author received an answer that worked for him or her personally, but not every user comes back to accept an answer, and of those who do, they may not change the accepted answer if a newer, better answer comes along later.''} \cite{sof_accepeted}

Moreover, in order to improve the quality of questions asked in the community, StackOverflow adopts some qualification methods. In particular, low-quality questions will be put on hold state if they don't suit the community. When a question is put on hold, it can not be answered but can be edited to get suited for answering \cite{neshati_IPM2017}. 
\section{Observations on data}
\label{sec:proof}
In this section, we discuss the necessity of translation models and show why they are important to retrieve more relevant results and accordingly improve the result of expertise retrieval task.
\begin{figure}[h]
	\centering
	\includegraphics[scale=0.39]{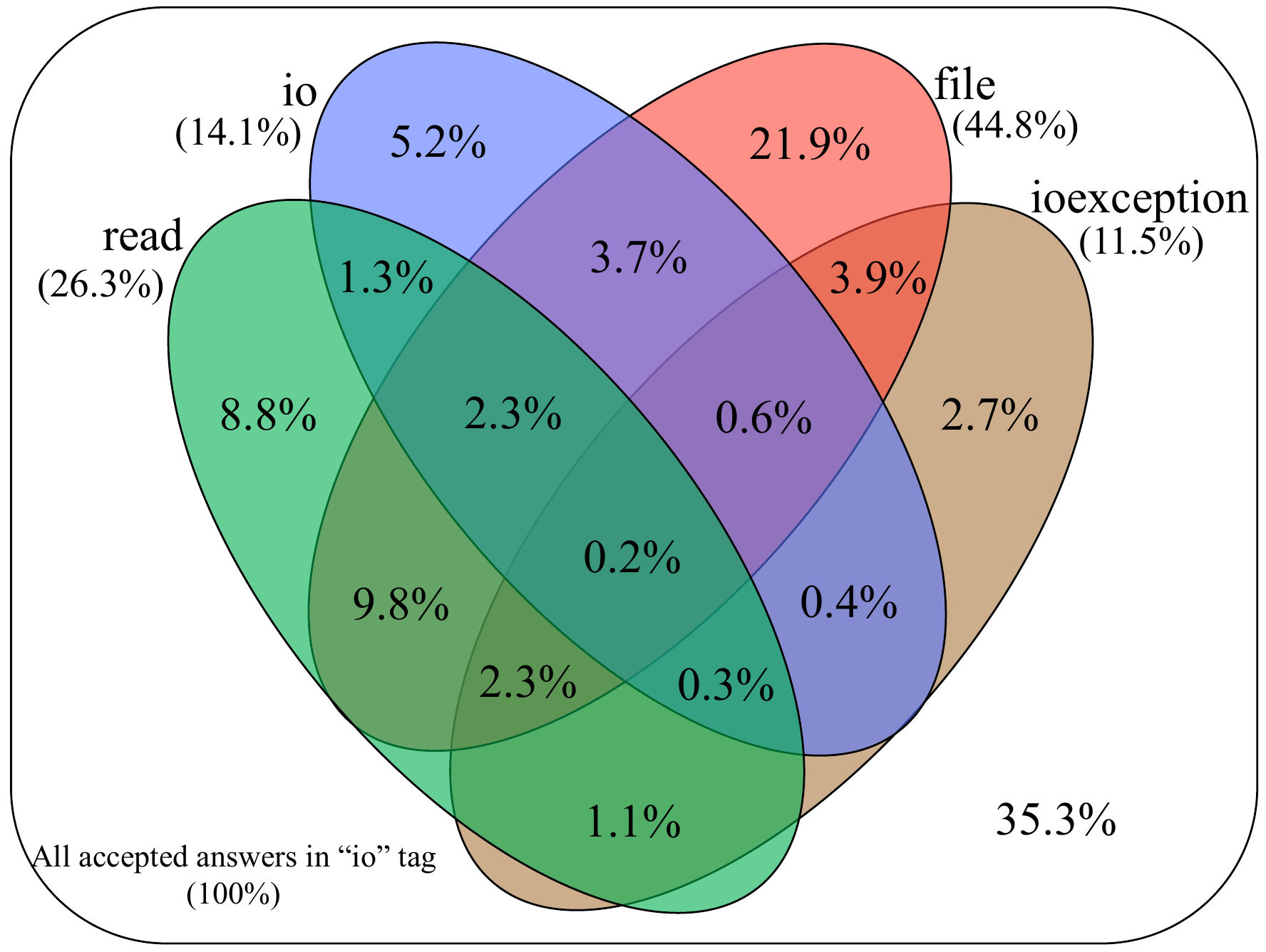}
	\caption{The Venn diagram of answers associated with questions tagged by ``io''}
	\label{io:b}
\end{figure}

\par

As mentioned in previous sections, a challenging problem of expert finding in StackOverflow is the vocabulary gap issue referring to the gap which exists between the main query and the terms used by experts. 
%Fig. \ref{io:b} represents this gap for the ``io'' query in StackOverflow dataset.
%As indicated, among the answers which are related to ``io'' (i.e. they have ``io'' as at least one of their tags), only 14.1\% of them include ``io'' in their body whereas, 44.8\% of these answers have ``file'' included in their body. This exemplifies the deep gap between users' information need (i.e. queries) and the textual representation of queries which can be bridged using translation terms.

Fig. \ref{io:b} represents a Venn diagram visualizing the occurrence of top four translation words (by MI translation approach described in section \ref{sec:mi_app}) among the answers which are related to ``io'' (i.e. they have ``io'' as at least one of their tags). These translations are ``file'', ``io'', ``read'' and ``ioexception''. As indicated, only 14.1\% of them include ``io'' in their bodies whereas, 64.7\% of them include at least one of the translations. To be more specific, 21.9\% of answers contain the term ``file'' and no other translations, 3.7\% include both ``file'' and ``io'' term, 0.6\% include ``file'', ``io'' and ``ioexception'' in their bodies and so forth. The answers including the term ``file'' are forming 44.8\% of all answers associated with ``io'' tag. Whereas, only 5.2\% of answers are covered by ``io'' term and no other translations. This exemplifies the deep gap between users’ information need (i.e. queries) and the textual representation of queries which can be bridged using translation terms.

\begin{figure}[]
	\centering
	\includegraphics[width=1\linewidth]{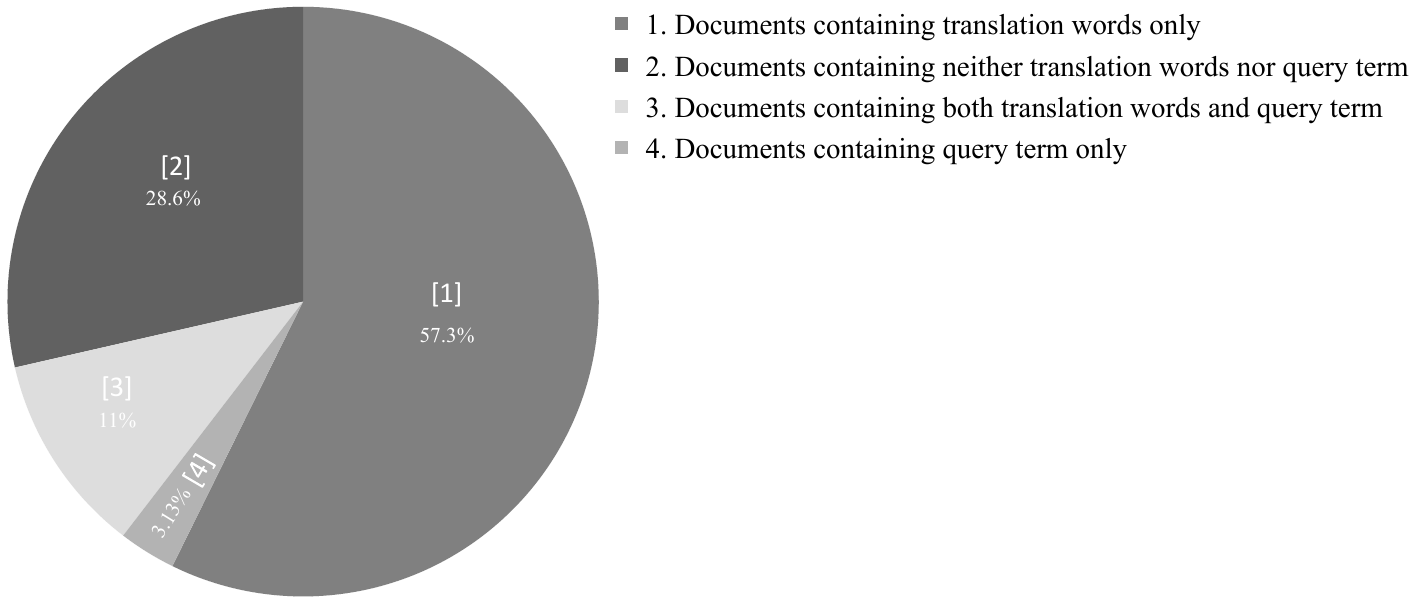}
	\caption{Share of ``io'' related documents retrieved by retrieval models}
	\label{io:a}
\end{figure}

%While Fig. \ref{io:b} indicates the vocabulary gap for three translations of ``io'' (i.e. ``file'', ``read'' and ``ioexception''), Fig. \ref{io:a} indicates the same information for 10 translations. According to this figure, surprisingly, 57.3\% of the ``io'' related answers have not included the term ``io'' itself in their body while covered at least one of the translation terms. This observation justifies the effectiveness of translation approach to improve the retrieval performance. 
%Moreover, two sectors (i.e. sectors 2 and 3) of the pie chart indicate the cases in which translation approaches cannot improve the retrieval performance directly. 
%Specifically, sector 2  indicates  28.6\% of the answers, related to ``io'' cannot be retrieved using neither ``io'' nor the related translations.  In other words, they cannot be retrieved using our proposed translation models at all. The third sector demonstrates the set of answers which include both the main query (i.e. ``io'') and at least one of the translation terms. The answers in this sector can be directly retrieved without using translation models. Finally, the last sector illustrates the answers which include only the main query (i.e. ``io'') and none of the top ten translations. Practically, the translation models cannot be used for this subset of answers, however, they form a very small portion of the whole related answers (i.e. only 3.1\%).
While Fig. \ref{io:b} indicates the vocabulary gap for three translations of ``io'' (i.e. ``file'', ``read'' and ``ioexception''), Fig. \ref{io:a} indicates the same information for 10 translations. According to this figure, surprisingly, 57.3\% of the ``io'' related answers have not included the term ``io'' itself in their body while covered at least one of the translation terms. This observation justifies the effectiveness of translation approach to improve the retrieval performance. 
Moreover, two sectors (i.e. sectors 2 and 3) of the pie chart indicate the cases in which translation approaches cannot improve the retrieval performance directly. 
Specifically, sector 2  indicates  28.6\% of the answers, related to ``io'' cannot be retrieved using neither ``io'' nor the related translations.  In other words, they cannot be retrieved using our proposed translation models at all. The third sector demonstrates the set of answers which include both the main query (i.e. ``io'') and at least one of the translation terms. The answers in this sector can be directly retrieved without using translation models. Finally, the last sector illustrates the answers which include only the main query (i.e. ``io'') and none of the top ten translations. Practically, the translation models cannot be used for this subset of answers, however, they form a very small portion of the whole related answers (i.e. only 3.1\%).

To sum up, both Figures \ref{io:a} and \ref{io:b} imply that the vocabulary gap in CQA networks, which exist between the user’s query and the terms which are used in the answers body, is remarkable. Thus, translation models should be utilized in order to improve expert finding. Each of these translations retrieves new documents which should be blended together to find the final ranking of the candidates. 
\par
It is worth mentioning that documents (i.e. answers) have not the equal quality in StackOverflow \cite{neshati_IPM2017}. Therefore, it is necessary to take the quality of answers into consideration in scoring step. A simple approach to measure the quality of the answers is \textit{Voteshare}. To be more specific, for every question asked in StackOverflow, a competition is formed among answerer to get more votes. It is obvious that better answers of a specific question receive a higher share of votes compared to the other answers. Since better answers are expected to be provided by expert users, it can be inferred intuitively that these answers have a higher Voteshare.
Voteshare, as it implies, is the share of an answer's votes to the summation of all answers' votes in a single thread as shown in Eq. \ref{eq:voteshare}.\footnote{We assumed that answers with zero/negative votes have not any Voteshare.}

\begin{equation}
Voteshare(a_i) = \frac{Vote(a_i)}{\displaystyle \sum_{j=1}^{j=n} Vote(a_j)}
\label{eq:voteshare} 
\end{equation}
where $n$ is the total number of answers in a thread.
\begin{figure}[h]
	\centering
	\includegraphics[width=0.8\linewidth]{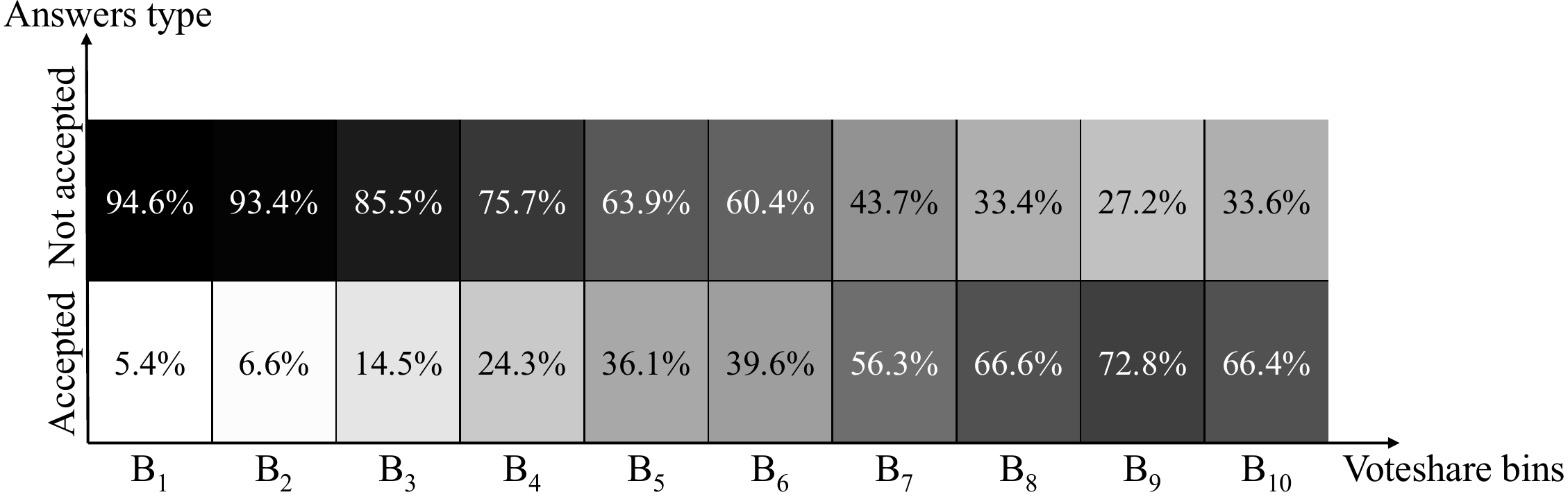}
	\caption{Distribution of Voteshare on high and low quality answers}
	\label{fig:voteshare}
\end{figure}
Fig. \ref{fig:voteshare} indicates the relation between Voteshare and the quality of answers. In this figure, the answers are clustered into 10 bins with regard to their Voteshare values. To be more specific, $B_1$ includes the decile of answers with lowest Voteshare values,  accordingly, $B_{10}$ is decile of answers having the highest value of Voteshare. Additionally, we categorized answers into two groups including accepted answers (i.e. high-quality answers) and not accepted answers \footnote{The concept of accepted answers is described in section \ref{sec:sof}.} (i.e. low-quality answers). According to Fig. \ref{fig:voteshare}, answers with slightly inferior value of Voteshare have lower chance of being accepted and accordingly they have less quality. In contrast, answers with superior value of Voteshare (e.g. Bin 8 and above) are most probably high-quality answers. For instance, 66.4\% of answers in $B_{10}$ are accepted, whereas, only 5.4\% of the answers are marked as accepted in $B_1$.
		
\par
In section \ref{sec:blending} we will propose four methods in order to score the candidates using the documents which are retrieved via translation models. Two of these approaches is based on the Voteshare concept.

\section{Baselines}
\label{sec:baselines}
In this section, we will explain our examined baselines in detail. First of all, we will review expert finding task mathematically. Then, we will explain probabilistic language models for expert finding. Finally, we will describe topic modeling approach which is among successful approaches to address the vocabulary gap issue.

In order to estimate $P(ca|q)$, different methods have been proposed \cite{balog:trends, Momtazi} to address this task. Prior to explaining the approaches which are taken as our baselines, it would be favorable to investigate the expert finding task and its nature. The task of expert finding is defined as identification and ranking of candidates who are expected to be expert with respect to a given query. Thus, in expert finding task, we tend to indicate $P(ca|q)$ and then rank the candidates with regard to this probability. Obviously, the higher a candidate has $P(ca|q)$, the more probable he is to be an expert candidate. $P(ca|q)$ can be approximated using Bayes' theorem as follows.

\begin{equation}
P(ca|q) = \frac{P(q|ca) P(ca)}{P(q)},
\end{equation}

in which, $P(ca)$ is the prior probability of candidate $ca$, $P(q)$ is the probability of query $q$ and can be ignored. As it has a constant value for a given query, it does not affect ranking of experts. Therefore, the probability of candidate given a query (i.e. $P(ca|q)$) is directly proportional to the probability of a query given the candidate $P(q|ca)$ and weighted by a prior probability of candidate.

\begin{equation}
P(ca|q) \propto P(q|ca) P(ca).
\end{equation}

\subsection{Language Models for Expert Finding} 
\label{sec:relwork_balog}
Balog \textit{et al.} \cite{balog2009} have proposed two generative probabilistic language models, known as \textit{candidate-based} and \textit{document-based} approaches. Each one models the expert finding from slightly distinctive perspective. They are defined as follows.

\subsubsection{Candidate-based approach}
   The first model, which is known as candidate-based model and referred as Model 1, approximates the corresponding probability (i.e.  $P(q|ca)$) from candidates' point of view. Indeed, it builds a multinomial language model $\theta_{ca}$ for each candidate over the terms which are used in their associated documents. Under the assumption that query terms are independently sampled, $P(q|ca)$ can be calculated by the production of terms of the query as follows.

\begin{equation}
\label{eq:balogM1_lm}
P(q|ca) = P(q|\theta_{ca}) = \prod_{t \in q}{P(t|\theta_{ca})}^{n(t,q)}
\end{equation}
where $n(t,q)$ is the number of times which term $t$ appears in query $q$. In order to estimate $P(t|\theta_{ca})$, firstly, the probability of a term given a candidate $P(t|ca)$ must be estimated. It should be noted that some candidates may not use a specific term of a query and thus make the probability equal to zero. Hence, in order to avoid zero probabilities due to data sparsity, $P(t|ca)$ should be smoothed with the background collection probabilities as shown in Eq.  \ref{eq:balogM1_lms}.
\begin{equation}
\label{eq:balogM1_lms}
P(t|\theta_{ca}) = (1-\lambda_{ca}) P(t|{ca}) +‌ \lambda_{ca} P(t)
\end{equation}
where $P(t)$ is the probability of a term in the documents collection, $P(t|ca)$ is the likelihood that candidate $ca$ would write about term $t$ and is estimated using Eq. \ref{eq:balogM1_ptca}, and $\lambda_e$ is the parameter of model. 

\begin{equation}
\label{eq:balogM1_ptca}
P(t|ca) = \sum_{d \in D_{ca}}{P(t|d, ca)}.{P(d|ca)}
\end{equation}

Assuming that document and the candidate are conditionally independent, $P(t|d, ca)$ can be reduced to $P(t|d)$ in which is the occurrence probability of term $t$ in document $d$. $P(t|d)$ can also be approximated using $P_{MLE}$ (i.e. maximum-likelihood probability). Candidate model is obtained by combining the Eqs. \ref{eq:balogM1_lm}-\ref{eq:balogM1_ptca} as shown in the following equation.

\begin{equation}
\label{eq:balogM1}
\begin{aligned}
& P(q|ca) = \\
& \displaystyle\prod_{t \in q}{}\Big\{(1-\lambda_{ca}).\Big(\displaystyle\sum_{d \in D_{ca}}{P(t|d).P(d|ca)}\Big) + \lambda_{ca} . P(t)\Big\}^{n(t,q)}
\end{aligned}
\end{equation}

\subsubsection{Document-based approach} \label{sec:baseline_balog_DM}
The second approach of expert finding adopted by Balog et al. acts somewhat differently to estimate $P(q|ca)$. It is known as document-centric model (referred as Model 2) and in spite of candidate-based model which found candidates directly, it considers the documents in a collection as a bridge which links the given query to candidates and evidence their author's expertise. In this case, the problem of expert finding can be defined as follows. Given a collection of documents which are ranked according to the given query, the authors of the relevant documents to the query should be retrieved and ranked. This model can be set off by taking the sum over entire documents $d \in D_{ca}$ as expressed in Eq. \ref{eq:balogM1_pte}.

\begin{equation}
\label{eq:balogM1_pte}
P(q|ca) = \sum_{d \in D_{ca}}{P(q|d,ca).P(d|{ca})}
\end{equation}
where $P(q|d,ca)$ is the likelihood of generating query $q$ according to document $d$ and candidate $ca$, and $P(d|ca)$ denotes the binary association of document $d$ and candidate $ca$. Under the assumption that query terms are occurred independently, $P(q|d,ca)$ can be estimated as follows.

\begin{equation}
\label{eq:balogM2_pqd}
P(q|d,{ca}) = \prod_{t \in q} P(t|d,{ca})^{n(t,q)}
\end{equation}
Having substituted Eq. \ref{eq:balogM2_pqd} into Eq. \ref{eq:balogM1_pte}, we result the following equation.

\begin{equation}
\label{eq:balogM2_pqe2}
P(q|ca) = \sum_{d \in D_{ca}}{\prod_{t \in q} P(t|d,ca)^{n(t,q)}.P(d|ca)}
\end{equation}

In order to estimate $P(t|d,ca)$, we can assume conditional independence between the query $q$ and the candidate $ca$ i.e. $P(t|d,ca) \approx P(t|\theta_{d})$ in which $\theta_{d}$ is the document language model which is inferred from document $d$. Therefore, the probability of a term $t$ given document model $\theta_{d}$ can be calculated using Eq. \ref{eq:balogM2_ptd}.

\begin{equation}
\label{eq:balogM2_ptd}
P(t|\theta_d) = (1 - \lambda_d).P(t|d) + \lambda_d.P(t)
\end{equation}

By putting $P(t|\theta_{d})$ instead of $P(t|d,ca)$ in Eq. \ref{eq:balogM2_pqe2}, final approximation of document-based model is yielded as follows.

\begin{equation}
\label{eq:balogM2}
\begin{aligned}
&P(q|ca) = \\ 
& \displaystyle\sum_{d \in D_{ca}}{\displaystyle\prod_{t \in q}{}\Big\{(1-\lambda_d) . P(t|d) + \lambda_d . P(t)\Big\}}^{n(t,q)}.P(d|{ca})
\end{aligned}
\end{equation}

\subsection{Topic Modeling for Expert Finding}
\label{sec:relwork_momtazi}
Topic modeling is the other baseline to determine the probability of a candidate to be an expert with regard to a given query (i.e. $P(ca|q)$). Momtazi and Naumann \cite{Momtazi} have proposed the model for the task of expert finding. The model approaches the document-based approach with this difference that it utilizes topics extracted from a document repository (i.e. collection) rather than documents. Indeed, the extracted topics are acted as a bridge to connect candidates to a given query. It is worth mentioning that topic modeling approach is among vigorous approaches to overcome the vocabulary gap issue and also is expected to outperform state-of-the-art approaches (i.e. candidate-based and document-based).
The process of expert finding using topic modeling includes two main phases. The first phase is involved with extracting topics using Latent Dirichlet Allocation
(LDA) and mostly performed off-line. In the next phase, the extracted topics are used to determine $P(q|ca)$ as follows.

\begin{equation}
\label{eq:tm_pqe1}
P(q|ca) = \sum_{z \in Z}{P(q|z,ca) P(z|ca)}
\end{equation}
in which $Z$ symbolizes the extracted topics, and $P(q|z,ca)$ is the likelihood of generating query $q$ given topic $z$ and candidate $ca$. Under the conditional independence assumption between $q$ and $ca$, $P(q|z,ca)$ can be reduced to $P(q|z)$. The probability of query given topic (i.e. $P(q|z)$) is calculated by taking the product over the terms of query $q$ as expressed in Eq. \ref{eq:tm_pqz}.

\begin{equation}
\label{eq:tm_pqz}
P(q|z) = \prod_{t \in q}{P(t|z)^{n(t,q)}}
\end{equation}
where $t$ denotes query term, and $n(t,q)$ is the number of times that $t$ appears in $q$.

We can also estimate the probability of topic $z$ given candidate $ca$ (i.e. $P(z|ca)$) using Bayes' theorem as follows.

 \begin{equation}
 \label{eq:tm_pqe2}
 P(z|ca) \propto P(ca|z) P(z)
 \end{equation}
in which $P(z)$ is the prior probability of selecting topic $z$ and generally it is considered to be uniform. Substituting Eqs. \ref{eq:tm_pqe2} and \ref{eq:tm_pqz} into Eq. \ref{eq:tm_pqe1} leads to the following equation.

\begin{equation}
\label{eq:tm_pqe3}
P(q|ca) = \sum_{z \in Z}{
	\Big [\prod_{t \in q}{P(t|z)^{n(t,q)}}\Big] P(ca|z) P(z)
}
\end{equation}
 where $P(ca|z)$ is the probability that topic $z$ would be talked by candidate $ca$ and calculated using LDA algorithm. To avoid zero probabilities, Jelinek-Mercer smoothing is employed. So in this way, a background probability is interpolated with the original probability to ensure that there are not any zero probability (the probability is always non-zero). By applying Jelinek-Mercer smoothing to Eq. \ref{eq:tm_pqe3}, the final estimation of topic modeling approach is resulted as follows.
 
 \begin{equation}
 \label{eq:tm_smooth}
\begin{aligned}
 &P(q|ca) = \\ 
 & \sum_{z \in Z}{
 	\prod_{t \in q}{  \Big \{
 		(1 - \lambda) P(t|z) +‌ \lambda P(t)
 		\Big  \} ^{n(t,q)} P(ca|z) P(z)
 	}	
 }
\end{aligned}
 \end{equation}
 
\section{Translation Approaches} \label{sec:approach_translation}
As mentioned in the previous section, the translation approach can be beneficial to reduce the gap between query and the terms occurred in documents. Here, each query represents a \textit{skill area} that can be used to retrieve relevant candidates. In the rest of this paper, we demonstrate the expert finding query by $sa$ notation\footnote{The $tag$, $sa$ and $q$ refer to the same concept in this paper.}. 
In this section, we explain two methods of skill area translation which are Mutual Information approach (i.e. MI) and Word Embedding approach (i.e. WE), respectively.

\subsection{Mutual Information Based Approach (MI)}
\label{sec:mi_app}
Assuming each skill area as a class label, the set of answers in StackOverflow can be partitioned into two disjoint subsets. The first subset contains answers tagged by a given skill area, and the second one includes other answers.
In our problem, we can use the mutual information (MI) to determine how much information the presence or absence of a term contributes to making the correct classification decision \cite{manning}. For each pair of word $w$ and skill area $sa$, the MI can be calculated using the following equation.

\begin{equation}
MI(sa,w) = \sum_{ A_{sa} = 0,1}{
	\sum_{A_{w} = 0,1}{
		p(A_{sa} , A_{w}) \log{\frac{p(A_{sa} , A_{w})}{p(A_{sa}) p(A_{w})}}
	}%inner sum
}%First sum
\label{eq:main_mi}
\end{equation}
in which $A_{sa}$ and $A_{w}$ denote binary variables indicating the occurrence event of skill area $sa$ and word $w$ in an answer. The probabilities indicated in Eq. \ref{eq:main_mi} can be estimated using the following equations: 

\begin{align*}
p(A_{sa} = 1)  &=  \frac{c(A_{sa} = 1)}{N} \\
p(A_{sa} = 0)  &=  1 - p(A_{sa} = 1) \\
p(A_{w} = 1)  &=  \frac{c(A_{w} = 1)}{N} \\
p(A_{w} = 0)  &=  1 - p(A_{w} = 1) \\
p(A_{sa} = 1 , A_{w} = 1)  &= \frac{c(A_{sa} = 1 , A_{w} = 1)}{N}\\
p(A_{sa} = 1 , A_{w} = 0)  &= \frac{c(A_{sa} = 1) - c(A_{sa} = 1 , A_{w} = 1)}{N}\\
p(A_{sa} = 0 , A_{w} = 1)  &= \frac{c(A_{w} = 1) - c(A_{sa} = 1 , A_{w} = 1)}{N} \\
p(A_{sa} = 0 , A_{w} = 0)  &= 1 - p(A_{sa} = 1 , A_{w} = 1) \\
- p(A_{sa} = 1 , &A_{w} = 0) - p(A_{sa} = 0 , A_{w} = 1)
\end{align*}
%\[ p(X_{t} = 1)	=	\frac{c(X_{t} = 1)}{N} \]
%\[ p(X_{t} = 0)	=	1 - p(X_{t} = 1) \]
%\[ p(X_{w} = 1)	=	\frac{c(X_{w} = 1)}{N} \]
%\[ p(X_{w} = 0)	=	1 - p(X_{w} = 1) \]
%\[ p(X_{t} = 1 , X_{w} = 1)  = \frac{c(X_{t} = 1 , X_{w} = 1)}{N} \]
%\[ p(X_{t} = 1 , X_{w} = 0)  = \frac{c(X_{t} = 1) - c(X_{t} = 1 , X_{w} = 1)}{N} \]
%\[ p(X_{t} = 0 , X_{w} = 1)  = \frac{c(X_{w} = 1) - c(X_{t} = 1 , X_{w} = 1)}{N} \]
%\[ p(X_{t} = 0 , X_{w} = 0)  = 1 - p(X_{t} = 1 , X_{w} = 1) \]  \[ - p(X_{t} = 1 , X_{w} = 0) - p(X_{t} = 0 , X_{w} = 1) \]
where $c(A_{sa} = 1)$ and $c(A_{w} = 1)$ imply number of the answers associated with skill area $sa$, and  denotes number of answers containing word $w$, respectively. $N$ is also the number of all answers. To obtain translation probability, the MI score should be normalized using Eq. \ref{eq:normalized_mi}. For a given skill area $sa$, the most enlightening words can be sorted using $p_{MI}(w|sa)$ probability as follows.

\begin{equation}
p_{MI}(w|sa) = \dfrac{MI(sa,w)}{\sum_{w'}{MI(sa,w')}}
\label{eq:normalized_mi}
\end{equation}
$p_{MI}(w|sa)$ gives the probability of translating skill area $sa$ to word $w$. Intuitively, this probability will be higher, if the word $w$ and skill area $sa$ are likely to co-occur with each other.

\begin{figure*}
	\centering
	\includegraphics[width=0.9\linewidth]{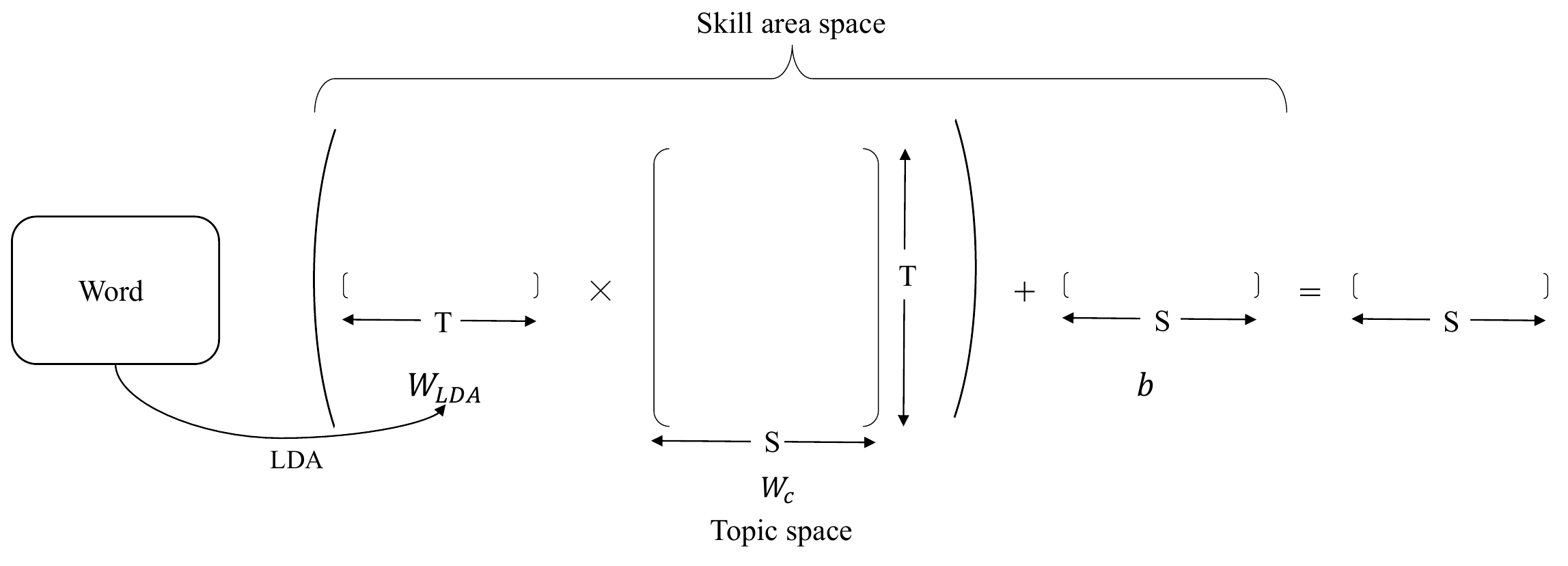}
	\caption{Schematic representation of  the proposed word embedding model}
	\label{fig:we_shape}
\end{figure*}

\subsection{Word Embedding Based Approach}
\label{sec:we_app}
As mentioned in section \ref{sec:relwork_momtazi}, the proposed model of Momtazi \textit{et al.} \cite{Momtazi} maps expert candidate, documents, and their used terms to a \textit{topic space} and the matching between them is formulated in the corresponding space.
Having reduced the vocabulary gap, the topic modeling approach improves the retrieval performance over the profile-based and document-based models introduced in section \ref{sec:relwork_balog}. Though, since terms representing skill areas (e.g. ``hibernate'', ``orm'' and etc.) rarely occurs in documents, it is necessary to embed document terms and skill areas i.e. queries into a single new space which is called \textit{skill area space}.

%\begin{enumerate}
%	\item Terms representing skill areas (e.g. ``hibernate'', ``orm'' and etc.) rarely occurs in documents.
%	\item A single skill area may be related to more than one topic extracted by topic modeling and conversely a topic may also be related to more than one single skill area. 
%\end{enumerate}

Having embedded skill areas and document terms into the same space, in this section, we proposed a domain-aware translation method which maps a given skill area to the most relevant words occurred in documents (i.e. answers of StackOverflow in our problem).
\par
As indicated in Fig. \ref{fig:we_shape}, we start translation process by applying topic modeling approach to the given set of documents to obtain a low-dimensional i.e. \textit{topic space} representation of each word in our dataset. Then, a mapping function is designed from the \textit{topics space} to the \textit{skill areas space}. Using this function, the words of documents and the tags, which represent skill areas, are embedded into a single low-dimensional space. For notational convenience, we write $P(\textbf{\textit{sa}} | .)$ for the probability distribution over skill areas, which is the result of vector arithmetic.
 The relevance probability of the skill areas given a word $w$ can be expressed by the following equation.

\begin{equation}
P_{WE} (\textbf{\textit{sa}}|w)=\frac{1}{z}e^{W_{LDA} .W_C+b} 
\label{embed}
\end{equation}
where $W_{LDA}$ is a $1\times T$ vector expressing word $w$ in topic space ($T$ is the number of topics), $W_C$ is a $T\times S$  matrix which maps the topic space representation of word $w$ to skill area space ($S$ symbolizes the number of skill areas), $b$ is a $1\times S$ vector which represents the prior relevance probability of skill areas to a given word, and finally $z$ is a normalization factor which is calculated as follows: $z = \sum_{j=1}^{|sa|}[e^{W_{LDA} .W_C+b}]_j$.

%In this model, the matrix $W_C$ and vector $b$ are denoting unknown parameters and should be learned during training. They are estimated using error back-propagation algorithm. During training, for a set of given documents with known tags i.e. skill areas, we estimate the observed occurrence probability of each word in the skill areas as follows.

In this model, the matrix $W_C$ and vector $b$ are denoting unknown parameters and should be learned during training. They are estimated using error back-propagation algorithm. During training step, for a set of given documents (i.e. training data) with known tags i.e. skill areas, we estimate the observed occurrence probability of each word in the skill areas as follows.\footnote{In real scenarios, most of the times, it is not possible to compute $P_{observed}$, accordingly, we estimated it observing only a subset of data (i.e. training data).}

%% W{lda} daraje azadi yadgiri ra kam karde va baes mishavad kalamati ke serfan dar yek skill area rokhdad daranad (masalan trac dar io ke ehtemale 1 darad) zarib paeen be an ekhtesas midahad ke ba zarb shodan dar addad haye bozorg meghdar ziadi peyda nemikonad ama dar avaz kalamti manande bufferedreader ke dar tedad ziadi topic (albate dar io ba tedad bisgatar) rokh dade and balatar miayand.
\begin{equation}
P_{observed} (\textbf{\textit{sa}}|w)=\frac{tf(\textbf{\textit{sa}},w)}{tf(w)}
\label{observed}
\end{equation}
in which $tf(sa,w)$ is the term frequency of $w$ in the documents which are tagged by $sa$, and $tf(w)$ denotes the term frequency of $w$ in training set. During model construction, we optimize the cross entropy of $H(P_{WE},P_{observed})$ using batch gradient descent as shown in Eq. \ref{eq:lf}.
\begin{equation}
\label{eq:lf}
L(W_{C},b) = \frac{1}{m} \sum_{i = 1}^{m} H(P_{WE},P_{observed})
+ \frac{\lambda}{2m}\left(  \sum_{i,j}{W_{C_{i,j}}^{2}} \right)
\end{equation} 
\begin{align*}
= - \frac{1}{m} \sum_{i = 1}^{m} { \left(
\sum_{j=1}^{|sa|} P_{observed}(sa_j|w_i) \log{P_{WE}(sa_j|w_i)}
\right) }
+ \frac{\lambda}{2m}\left(  \sum_{i,j}{W_{C_{i,j}}^{2}} \right)
\end{align*}

where $L(W_{C},b) $ gives us the loss function, $m$ is the size of a training batch, and $\lambda$ is a weight regularization parameter. The update rule for a particular parameter $\theta (W_C,b)$ given a single batch of size m is:
	\begin{equation}
	\label{eq:update}
\theta^{(t+a)} = \theta^{(t)} - \alpha^{(t)}  \odot \frac{\partial L(W_C^{(t)} , b^{(t)})}{\partial \theta}
	\end{equation} 

\begin{figure}[h]
	\centering
	\includegraphics[width=0.8\linewidth]{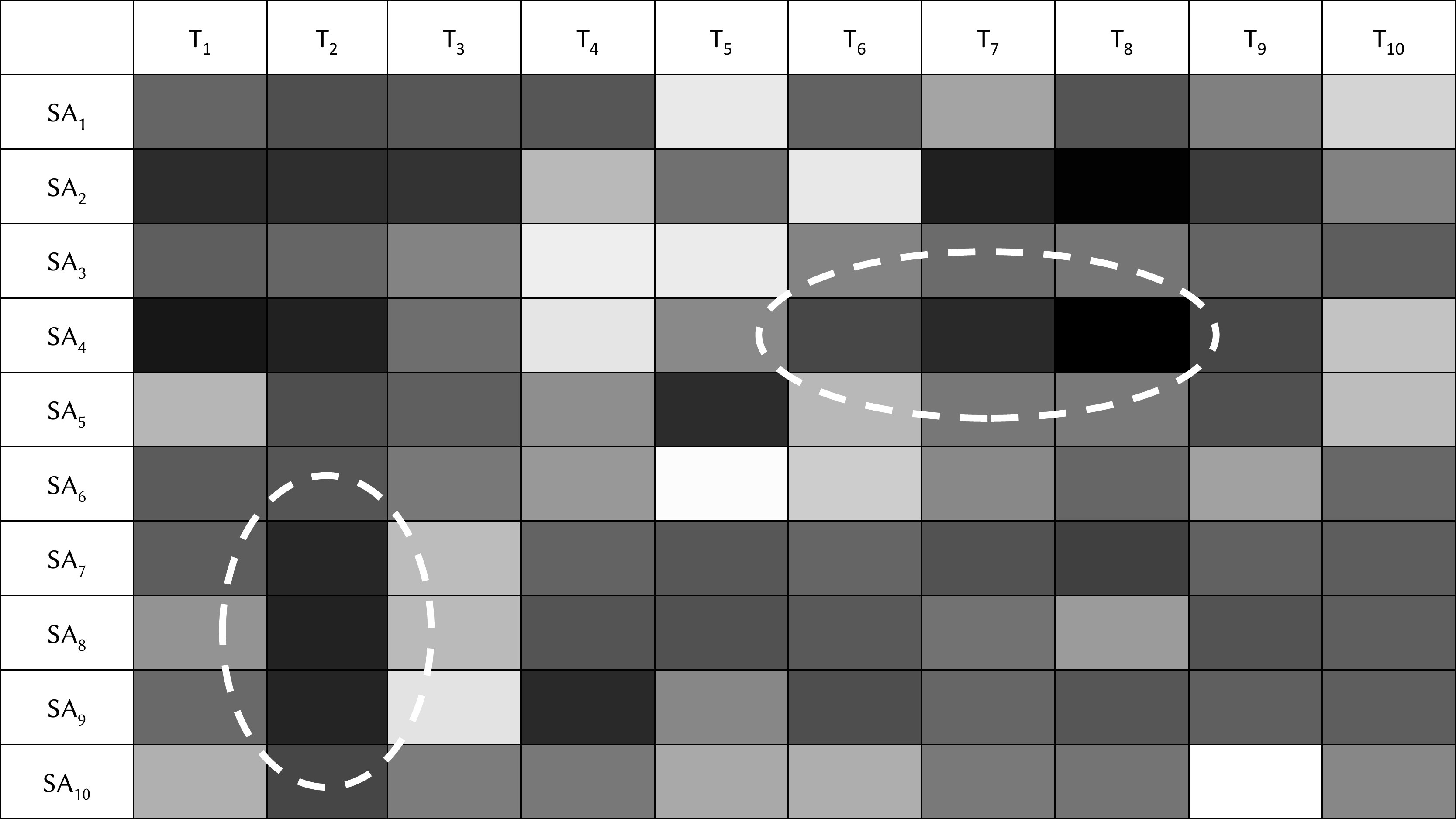}
	\caption{The heat-map of a subset of trained matrix
		$W_c$}
	\label{fig:heatmap}
\end{figure}

Suppose that word $w$ is related to topic $t_i$ and $t_j$ and numerously appeared in the answers associated with the skill areas $sa_k$ and $sa_m$. During training, the weights of the matrix $W_C$ and vector $b$ will be updated such that the representation of word $w$ in skill areas space be placed near the representation of $sa_k$ and $sa_m$. In addition, by applying the update rule, other words which are related to $t_i$ and $t_j$ will also get closer to $s_k$ and $s_m$ in skill areas space.

As pointed out before, the matrix $W_C$ gives a mapping function from topics space to skill areas space. Fig. \ref{fig:heatmap} depicts the heat-map of a subset of the matrix after training. The darker a cell is, the stronger association would be between the specific topic and skill area and vice versa. For example, skill area $SA_4$ is more associated with topics $T_7$, $T_8$ and $T_9$. While, $T_2$ is more associated with skill areas $SA_7$, $SA_8$ and $SA_9$. This figure illustrates a many-to-many relationship between topics and skill areas.

After the training process of matrix $W_C$ and vector $b$, $p(sa|w)$ can be estimated for each pair of words $w$ and skill area $sa$ using Eq. \ref{embed}. In order to find the most relevant translation terms for a given skill area $sa$, we use Bayes' theorem to estimate $p(w|sa) \approx p(w) p(sa|w)$, in which $p(w)$ denotes the prior probability of word $w$ to be chosen as a robust translation. We estimate $p(w)$ using TF-IDF computed over the data collection in our experiments.
\section{Scoring Approach} \label{sec:blending}
In this section, we propose our scoring approaches to estimate the final ranking of the candidates according to the translation terms extracted by the MI and WE methods. Suppose a given skill area $sa$ is translated to $w_1, w_2, ... , w_n$, which includes the skill area word as well (i.e. self translation). In expert finding problem, the goal is to estimate the probability of $P(ca|sa)$, by assuming the prior probability of candidates (i.e. $P(ca)$) to be uniform and having in mind that $P(sa)$ can be ignored in ranking, we have $P(ca|sa) \approx P(sa|ca)$, which is estimated as shown in Eq. \ref{eq:blend_tr_estimate}.

\begin{equation}
\label{eq:blend_tr_estimate}
P(sa|ca)  \overset{translate}{=\joinrel=} P(w_1, ... , w_n|ca)
\end{equation}
Each document $d$ associated with candidate $ca$ and contains one or more translation words can be considered as an evidence of the author's expertise on skill area $sa$. Therefore, following the idea of Model 2 proposed in \cite{balog2009}, the corresponding probability can be estimated as follows.

\begin{equation}
\label{eq:blend_balog2}
P(w_1, ... , w_n | ca) \propto \sum_{d \in D_{ca}}{P(w_1, ... , w_n | d,ca).P(d|ca)}
\end{equation}
in which $D_{ca}$ indicates the set of documents associated with candidate $ca$. By applying Bayes' theorem, we have $P(d|ca) = \frac{P(ca|d) . P(d)}{P(ca)}$. Since, in StackOverflow, each document (i.e. answer) has exactly one author. Therefore, we can assume that $P(ca| d) = 1$. Besides, $P(ca)$ is also assumed to be uniform. 
In addition, assuming conditional independence between the words and the candidate, the probability of $P(w_1, ... , w_n | ca)$ can be rewritten as follows.

\begin{equation}
\label{eq:blend_pw_new}
P(w_1, ... , w_n | ca) \propto \sum_{d \in D_{ca}}{P(w_1, ... , w_n | d) . P(d) }
\end{equation}
Where $P(w_1, ... , w_n | d)$ is \textit{document relevancy score} and $P(d)$ is \textit{document quality score} (prior probability of the document) of the documents. In this research, we have proposed two approaches for estimating each score which is described in the rest of this section.

\subsection{Estimating document relevancy score}
We have proposed two approaches to estimate $P(w_1, ... , w_n | d)$. 
In the first approach, we exploit the idea of language model to score each candidate. Following the Eq. \ref{eq:balogM2} in section \ref{sec:baseline_balog_DM}, the mentioned probability can be estimated as follows:
\begin{equation}
\label{eq:blend_lm_main}
P(w_1, ... , w_n | d) \propto \displaystyle\prod_{w_i}{}\Big\{(1-\lambda_d) . P(w_i|d) + \lambda_d . P(w_i)\Big\}
\end{equation}

Where, $P(w_i|d)$ is calculated by maximum likelihood estimation and $P(w_i)$    indicates the collection probability of word $w_i$. We refer this approach as \textit{Language Model Scoring} in the rest of this paper.

In the second approach to estimate $P(w_1, ... , w_n | d)$, instead of applying a probabilistic model (i.e. language model), we have focused on the number of expertise evidence occurred in each candidate's profile. Accordingly,  we consider each document $d$ associated with candidate $ca$ which contains one or more translation words as an evidence of the author's expertise on skill area $sa$. This is formally demonstrated in Eq. \ref{eq:blend_binary}. We refer this approach as \textit{Binary Scoring} in the rest of this paper.

\begin{equation}
\label{eq:blend_binary}
P(w_1, ... , w_n | d) = 
\left\{
\begin{array}{ll}
0, & \text{if } w_1, ... , w_n \notin \space d \\
1, & \text{otherwise}
\end{array}
\right.
\end{equation}

\subsection{Estimating document quality score}
In this section, we propose two estimation methods for document quality score (i.e. $P(d)$). In the first approach, we simply assume all documents have the same quality and accordingly, $P(d)$ can be ignored in ranking process. As a result Eq. \ref{eq:blend_pw_new} can be rewritten as follows:

\begin{equation}
\label{eq:blend_score_basic}
P(w_1, ... , w_n | ca) \propto \sum_{d \in D_{ca}}{P(w_1, ... , w_n | d)}
\end{equation}

In the second method, in order to take the quality of documents into account, utilizing the concept of Voteshare introduced in section \ref{sec:proof}, it is possible to estimate $P(d)$ by the Voteshare of document $d$. Accordingly, Eq. \ref{eq:blend_pw_new} can be rewritten as shown in the Eq. \ref{eq:blend_score_vs}. We refer this scoring approach as Voteshare based scoring in the rest parts of this paper.

\begin{equation}
\label{eq:blend_score_vs}
P(w_1, ... , w_n | ca) \propto \sum_{d \in D_{ca}}{P(w_1, ... , w_n | d) . Voteshare(d)}
\end{equation}

In both aforementioned equations, $P(w_1, ... , w_n | d)$ can be estimated by either language model or binary scoring methods which will lead to four disparate approaches to score the candidates.
\section{Experiments}
In this section, a set of experiments are designed to address the following research questions:

\begin{itemize}
	\item
	\textbf{RQ1: }Which models are more successful to overcome the vocabulary gap problem?
	\item
	\textbf{RQ2: }How the proposed scoring approaches can affect the overall performance of retrieval?
	\item
	\textbf{RQ3: }What is the effect of Voteshare in scoring step?
	\item
	\textbf{RQ4: }How many translations are enough to cover the vocabulary gap? How sensitive are the proposed approaches on the number of translations?
	\item
	\textbf{RQ5: }Is there any difference between translation provided by Mutual Information and Word Embedding approaches?
	
\end{itemize}
In the rest of this section, we first set forth the experimental setup and parameter setting and then present our experimental results to answer the aforementioned research questions.

\subsection{Experimental Setup}
In this section, we describe our datasets and parameter setting.

\subsubsection{Data Collection}
Our dataset is downloaded from StackOverflow\footnote{https://archive.org/details/stackexchange}. It covers the period August 2008 until March 2015 and contains 24,120,523 posts.

\setlength{\abovetopsep}{0pt}
\setlength{\aboverulesep}{0pt}
\setlength{\belowrulesep}{0pt}
\setlength{\belowbottomsep}{0pt}

\begin{table}[htbp]
	\centering
	\small
	\caption{General statistics for ``Java'' and ``PHP'' data collections }
	\begin{tabular}{|c|c|c|c|c|}
		\hline
		DataSet & \#Q   & \#A   & \#C   & Avg Q-Rel \\
		\midrule
		Java  &       810,071  &    1,510,812  &       206,397  & 44.75 \\
		\midrule
		PHP   &       714,476  &    1,298,107  &       191,060  & 91.64 \\
		\hline
	\end{tabular}%
	\label{tab:dataset}%
\end{table}%

We have selected questions and their associated answers tagged by ``Java'' and ``PHP'' as two separated data collection in our experiments. The statistic related to each data collection including number of questions (\#Q), number of answers (\#A), number of candidates (\#C) and the average number of relevant candidates per query (Avg Q-Rel) is indicated in table \ref{tab:dataset}. 

We mark users as experts on each tag (i.e. skill area) when two conditions are met. First, similar to the definition proposed in \cite{derijke:2015:early}, the candidates should have ten or more\footnote{Since the number of question and answers in ``PHP'' dataset is less than ``Java'', we used 8 accepted answers as the threshold.} of their answers marked as accepted by the questioner. Second, following the idea proposed in \cite{sparrow2014}, the acceptance ratio of their answers should be higher than the average acceptance ratio (i.e. 40\%) in test collection. The first condition filters users with a low level of engagement and the second condition filters low-quality users.
Moreover, we select 100 top most frequent tags which co-occurred with ``Java" and ``PHP'' tags in our datasets as expert finding queries.
Our queries and implementations for all approaches including baselines is publicly available\footnote{http://tiny.cc/sofef}.

\subsubsection{Parameter Setting and Implementation Detail}
In our baseline models described in section \ref{sec:baselines}, we use $\lambda = 0.5$ as the smoothing parameter for both models in section \ref{sec:relwork_balog}. For topic modeling approach described in section \ref{sec:relwork_momtazi}, we tried different settings and finally set the number of topics to be 100.

We translate each skill area to top 10 most relevant words using the MI and WE methods. In WE method, we restrict the size of vocabulary to top $2^{16}$ most frequent words. 
In order to optimize the loss function (i.e. Eq. \ref{eq:lf}), we use adadelta ($\rho = 0.95$, $\epsilon = 10^{-6}$) \cite{adadelta} with batch gradient descent and weight decay $\lambda = 0.01$. 

Additionally, Tensorflow\footnote{https://www.tensorflow.org} has been used to calculate matrix operations on a Nvidia Titan X GPU. In order to index and search data and also implement the baselines and the proposed models, we have used Apache Lucene\footnote{https://lucene.apache.org/}.

\subsection{Experimental Results}
In this section, we aim to answer the research questions mentioned earlier. Our first experiment is carried out to finding an answer to \textbf{RQ1}: Which models are successful to overcome the vocabulary gap problem? Our next experiment is concerned with analyzing proposed scoring methods to answer \textbf{RQ2: }How the proposed scoring approaches can affect the overall performance of retrieval?
 A comparison between Voteshare and basic scoring approach is done in order to answer the next research question \textbf{RQ3}: What is the effect of Voteshare in scoring step? The next experiment is performed in order to determine the effect of increasing translation counts on the results which sets forth the \textbf{RQ4}: How many translations are enough to cover the vocabulary gap? How sensitive are the proposed approaches on the number of translations? Finally, our last experiment is a comparison between the translations provided by our two proposed models and accounts for the \textbf{RQ5}: Is there any difference between translation provided by Mutual Information and Word Embedding approaches?

\subsubsection{Analyzing the performance of models to overcome vocabulary gap problem}

\begin{table}[htbp]
	\centering
%	\scriptsize
	\caption{Comparison of the proposed models with baselines. Binary scoring results are reported.}
\scalebox{0.75}
{
	\begin{tabular}{|c|c|c|c|c|c|}
		\hline
		& \textbf{Method} & \textbf{MAP} & \textbf{P@1} & \textbf{P@5} & \textbf{P@10} \\
		\midrule
		\midrule
		\multirow{7.5}[6]{*}{\begin{sideways}\textbf{Java}\end{sideways}} & LM 1  & 0.377  & 0.560  & 0.500  & 0.440 \\
		& LM 2  & 0.362  & 0.540  & 0.482  & 0.425 \\
		& TM    & 0.434  & 0.550  & 0.530  & 0.488 \\
		\cmidrule{2-6}          & MI    & 0.478  & \textbf{0.660} & 0.604  & 0.529 \\
		& $\Delta$ LM1 & 26.8\% * & 17.9\% * & 20.8\% * & 20.2\% * \\
		& $\Delta$ TM & 10.1\% * & 20.0\% * & 14.0\% * & 8.4\% \\
		\cmidrule{2-6}          & WE    & \textbf{0.496} & 0.650  & \textbf{0.626} & \textbf{0.540} \\
		& $\Delta$ LM1 & 31.6\% * & 16.1\% * & 25.2\% * & 22.7\% * \\
		& $\Delta$ TM & 14.3\% * & 18.2\% * & 18.1\% * & 10.7\% * \\
		\midrule
		\midrule
		\multirow{8}[6]{*}{\begin{sideways}\textbf{PHP}\end{sideways}} & LM 1  & 0.335  & 0.570  & 0.524  & 0.479 \\
		& LM 2  & 0.309  & 0.520  & 0.478  & 0.437 \\
		& TM    & 0.401  & 0.530  & 0.550  & 0.491 \\
		\cmidrule{2-6}          & MI    & 0.458  & 0.590  & 0.612  & 0.561 \\
		& $\Delta$ LM1 & 36.7\% * & 3.5\% & 16.8\% * & 17.1\% * \\
		& $\Delta$ TM & 14.3\% * & 11.3\% * & 11.3\% * & 14.3\% * \\
		\cmidrule{2-6}          & WE    & \textbf{0.509} & \textbf{0.600} & \textbf{0.626} & \textbf{0.581} \\
		& $\Delta$ LM1 & 52.0\% * & 5.3\% & 19.5\% * & 21.3\% * \\
		& $\Delta$ TM & 27.0\% * & 13.2\% * & 13.8\% * & 18.3\% * \\
		\hline
		
		\addlinespace[1ex]
		\multicolumn{6}{l}{ 
			\textsuperscript{*}{\textit{indicates that improvement is statistically significant}}} \\
		\multicolumn{6}{l}{ \textit{on a two-tailed paired t-test ($\alpha = 0.05$)}}
	\end{tabular}%
}%scalebox
	\label{tab:rq1}%
\end{table}%

Table \ref{tab:rq1} indicates the result of Language Model (LM) 1 and 2, Topic Modeling (TM), Mutual Information (MI) and Word Embedding (WE) scored by the basic binary approach. According to this table, two main observations can be concluded. First, the performance of TM is significantly better than LM approaches. This observation indicates that TM approach can overcome the vocabulary gap problem to some extents. Second, the MI and WE approaches even with basic scoring approach outperform TM. As mentioned before, a skill area can be mapped to more than one topic and each conversely a topic can be mapped to more than one skill area in our problem. However, in the TM approach, each answer is mapped to a few number of topics but the relationship between skill areas and answers is not determined directly. In contrast, the MI and WE methods directly extract top relevant translations for a given skill area and accordingly they surpass in terms of both precision and recall. As a result, these methods improve precision@n (P@n) and mean average precision (MAP) measures significantly in comparison with TM method. This observation is consistent on both ``Java'' and ``PHP'' datasets. Another interesting observation is that the TM method decreases P@1 measure on both data collections which means although this method is successful in reducing the vocabulary gap, and accordingly improving the recall, it slightly decreases the precision measure in top level rankings. In contrast, our proposed models not only increase the recall but also marginally improve the precision on both data collections.

\subsubsection{Analyzing performance of scoring methods}

\begin{table}[htbp]
  \centering
  \caption{Comparison of  scoring methods with baselines based on the MAP measure}
    \begin{tabular}{|l|cc|}
    \hline
    Method & Java &  PHP \\
    \midrule
    Language Model 1 & 0.377 & 0.335 \\
    Language Model 2 & 0.362 & 0.309 \\
    Topic Modeling & 0.434 & 0.401 \\
    \midrule
    Basic Language Model & 0.371 & 0.341 \\
    Voteshare based Language Model & 0.419 & 0.366 \\
    \midrule
    Basic Binary Scoring & 0.496 & 0.509 \\
    Voteshare Based Binary Scoring & 0.660 & 0.562 \\
    \hline
    \end{tabular}%
  \label{tab:rq1.5}%
\end{table}%

The retrieval performance of baselines and scoring methods on WE translations are demonstrated in Table \ref{tab:rq1.5}. As indicated in this table,  performance of Language Model based Scoring (LMS) are significantly less than binary scoring approaches. In some cases, LMS methods' performance is even below the baseline methods. This observation can be illustrated by two explanations. 
First, In binary scoring methods each document containing at least one translation method is assumed as a single vote of expertise evidence. Whereas, In LMS approaches, abundance occurrence of a translation term in a document will lead to large score for its owner and significantly affect the amount of evidence for that document. However, publishing a single post including a large number of relevant terms cannot essentially indicates expertise of a candidate, accordingly, it should not over-affect the score of each candidate. 
Take, ``io'' skill area as an example, it could be translated into ``stream'', ``file'', etc. Since programming codes form up a large portion of posts in StackOverflow and ``stream'' is commonly used in programming codes, it is occurred in a large number of documents. Consequently, superior Term Frequency (TF) of this word would unfairly increase the score of these documents as expertise evidence. In other words, one reason behind the poor performance of LMS methods is considerable effect of TF on candidates scoring.
Second, although expanding query (i.e. skill area) in LMS methods would marginally increase the recall of performance, the precision is significantly decreased as a result of concept drift \cite{semantic} which will lead to inferior retrieval performance in comparison with other scoring approaches. Consequently, we will focus on binary scoring method in the next experiments.

\subsubsection{Voteshare vs. Basic scoring approach}

As explained before, the MI and WE with basic scoring approach (i.e. without including Voteshare) improve the recall measure  considerably (i.e. they can solve the vocabulary gap problem) in comparison with TM and LM models. In addition, they can slightly improve the precision of retrieval. As mentioned in section \ref{sec:proof}, all documents (i.e. answers) in StackOverflow do not have the same quality. The Voteshare scoring approach aims to solve this problem by exploiting high-quality answers in scoring step. Table \ref{tab:rq2} indicates the comparison of MI and WE translation models on basic and Voteshare based binary scoring approaches.
\begin{table}[htbp]
	\centering
	%	\small
	\caption{Comparison of scoring approaches for both MI and WE models. BS and VS indicate basic and Voteshare based binary scoring  approaches respectively. }
	\scalebox{1}
	{
		\begin{tabular}{|c|c|c|c|c|c|}
			\hline
			& \textbf{Method} & \textbf{MAP} & \textbf{P@1} & \textbf{P@5} & \textbf{P@10} \\
			\midrule
			\midrule
			\multirow{4.5}[6]{*}{\begin{sideways}\textbf{Java}\end{sideways}} & MI (BS) & 0.478  & 0.660  & 0.604  & 0.529 \\
			& WE (BS) & 0.496  & 0.650  & 0.626  & 0.540 \\
			\cmidrule{2-6}          & MI (VS) & 0.647  & 0.850  & \textbf{0.736} & 0.652 \\
			& $\Delta$ MI (BS) & 35.3\% * & 28.8\% * & 21.9\% * & 23.3\% * \\
			\cmidrule{2-6}          & WE (VS) & \textbf{0.660} & \textbf{0.860} & 0.728  & \textbf{0.661} \\
			& $\Delta$ WE (BS) & 33.1\% * & 32.3\% * & 16.3\% * & 22.4\% * \\
			\midrule
			\midrule
			\multirow{5}[5]{*}{\begin{sideways}\textbf{PHP}\end{sideways}} & MI (BS) & 0.458  & 0.590  & 0.612  & 0.561 \\
			& WE (BS) & 0.509  & 0.600  & 0.626  & 0.581 \\
			\cmidrule{2-6}          & MI (VS) & \textbf{0.587} & \textbf{0.750} & \textbf{0.726} & \textbf{0.642} \\
			& $\Delta$ MI (BS) & 28.3\% * & 27.1\% * & 18.6\% * & 14.4\% * \\
			\cmidrule{2-6}          & WE (VS) & 0.562  & \textbf{0.750} & 0.696  & 0.621 \\
			& $\Delta$ WE (BS) & 10.4\% * & 25.0\% * & 11.2\% * & 6.9\% \\
			\hline
			
			\addlinespace[1ex]
			\multicolumn{6}{l}{ 
				\textsuperscript{*}{\textit{indicates that improvement is statistically significant based}}}\\
			\multicolumn{6}{l}{ \textit{on a two-tailed paired t-test ($\alpha = 0.05$)}}
		\end{tabular}%
	}%scalebox
	\label{tab:rq2}%
\end{table}%
In this table, MI (BS) and WE (BS) indicate corresponding translation models with Basic binary scoring (BS) approach. MI (VS), and WE (VS) indicate translation models with Voteshare based binary scoring approach (VS). Three important observations are noticed here: First, Voteshare scoring approach remarkably improve the precision at top levels of ranking independent of the translation model and the dataset. Second, in the majority of measures, the performance of WE method with basic scoring approach outperforms the MI model with the same scoring. The only exception here is P@1 on Java data collection which MI performs better than WE but not noticeably. Third, the performance of WE and MI method with Voteshare scoring approach is almost the same on both test collections. This observation indicates that although WE and MI‌ approaches overcome the vocabulary gap problem with different mechanisms, the Voteshare scoring method - Utilizing only high-quality documents -  provides a consistent performance on both test collections independent of underlying translation models.

\subsubsection{Sensitivity analysis on number of translations}

\begin{figure*}[t!]
	\centering
	\begin{subfigure}[b]{0.5\textwidth}
		\centering
		\includegraphics[width=1\linewidth]{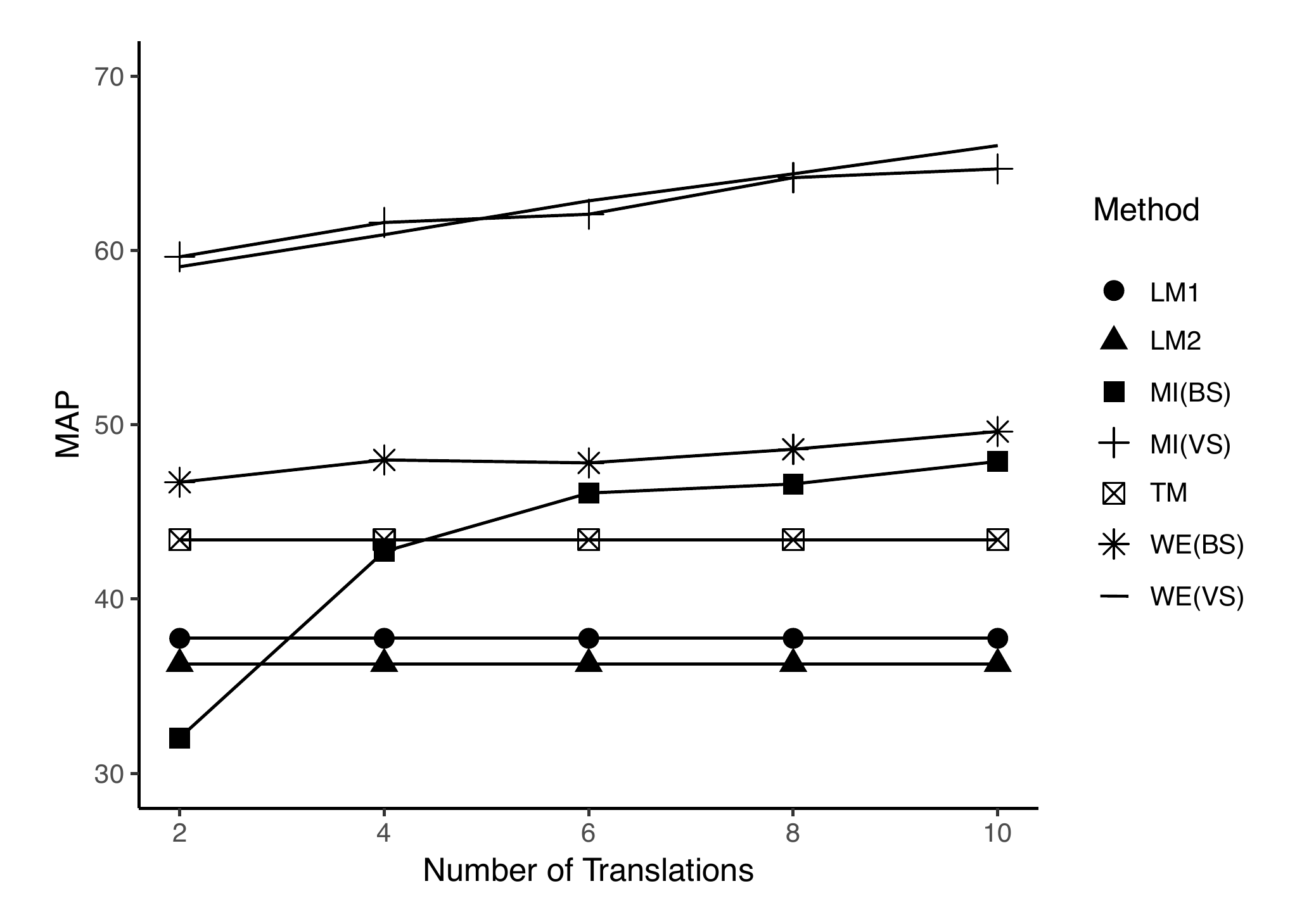}
		\caption{Java dataset}
		\label{subfig:counts_java}
	\end{subfigure}%
	~ 
	\begin{subfigure}[b]{0.5\textwidth}
		\centering
		\includegraphics[width=1\linewidth]{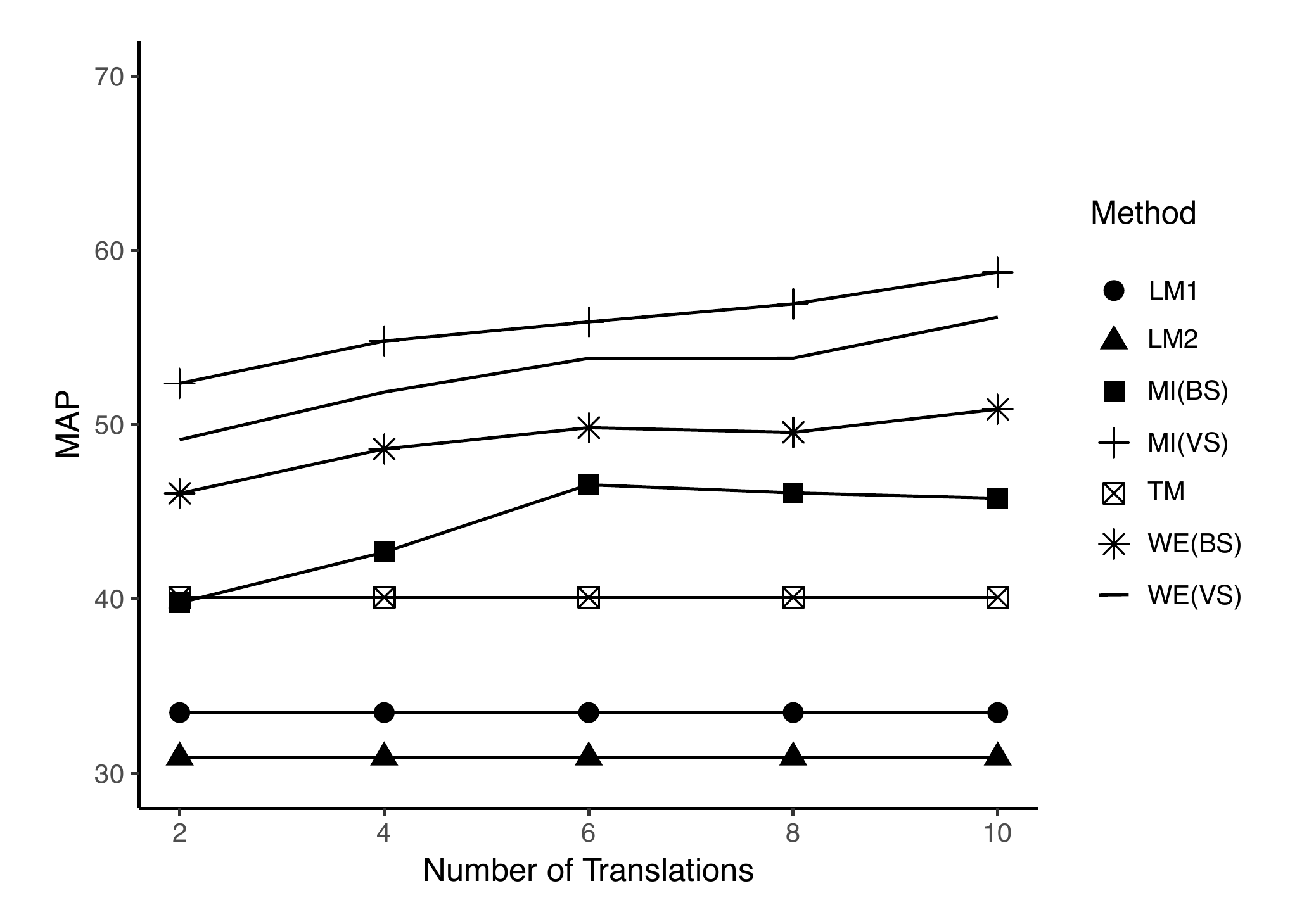}
		\caption{PHP dataset}
		\label{subfig:counts_php}
	\end{subfigure}
	\caption{The effect of varying number of translations on MAP measure for all proposed models}
	\label{fig:tr_count}
\end{figure*}

In this section, the proposed models are compared with the baselines in terms of number of translations for each original query which is the main parameter of the WE and MI models. Fig. \ref{subfig:counts_java} and Fig. \ref{subfig:counts_php} indicate the sensitivity of the proposed methods on the number of translation for ``Java'' and ``PHP'' data collections, respectively. Three important observations are marked here: First, the performance of WE and MI methods are almost ascending on both datasets independent of scoring approach. However, the performance of the proposed models does not change significantly after six translations. This observation is important because it means that the proposed models can be practically used in a retrieval engine without a significant overhead. Second, apparently, the MI method (with basic scoring) is more sensitive to the number of translation. In contrast, the WE method has a consistent performance even with only two translations on both test collections. Third, the translation models with Voteshare based scoring have almost the same performance (especially in ``Java'' test collection). This observation means that the Voteshare approach can work very well on top of both translation models. 

\subsubsection{Comparison of word embedding and mutual information translations}
\begin{table*}[htbp]
	\centering
	\caption{Sample skill area translations using word embedding and mutual information methods}

\scalebox{0.55}
{
	\begin{tabular}{|c|c|c|c|c||c|c|c|c|}
		\toprule
		\multirow{2}[4]{*}{\textbf{Method}} & \multicolumn{4}{c||}{Java}    & \multicolumn{4}{c|}{PHP} \\
		\cmidrule{2-9}          & \textit{Skill Area} & \multicolumn{1}{c|}{Translation 1} & \multicolumn{1}{c|}{Translation 2} & Translation 3 & \textit{Skill Area} & \multicolumn{1}{c|}{Translation 1} & \multicolumn{1}{c|}{Translation 2} & Translation 3 \\
		\midrule
		\multirow{3}[1]{*}{\textbf{MI}} & \textit{hibernate} & hibernate & entity & table & \textit{SOAP}  & xsd:string & s:element & soap:body \\
		& \textit{swing} & textsample & jframe & jpanel & \textit{GD}    & gd    & image & imagecreatetruecolor \\
		& \textit{selenium} & \_\_method.apply & selenium & webdriver & \textit{JSON}  & \multicolumn{1}{c|}{json} & \multicolumn{1}{c|}{json\_encode} & \multicolumn{1}{c|}{json\_decode} \\
		\midrule
		\multirow{3}[1]{*}{\textbf{WE}} & \textit{hibernate} & hibernate & entity & employee & \textit{SOAP}  & soap  & wsdl  & soapclient \\
		& \textit{swing} & jpanel & jbutton & jlabel & \textit{GD}    & image & img   & alt \\
		& \textit{selenium} & tests & junit & test  & \textit{JSON}  & json  & data  & json\_encode \\
		\bottomrule
	\end{tabular}%
}%scalebox
	\label{tab:rq4}%
\end{table*}%

Table \ref{tab:rq4} indicates the top three translations for a few number of skill areas extracted by MI and WE models on both datasets. Interestingly, the MI model usually translates the given skill area to more specific words while WE model selects more general words for the same topic. It seems that the MI model, which is basically a statistical translation model, is more sensitive to the co-occurrence of words and skill areas in documents.
As a result, the MI model in most cases selects pieces of program codes (e.g. ``\_\_method.apply'' for ``selenium'') which are the most frequent words found in StackOverflow answers. On the other hand, the WE model, as a semantic-aware translation model, provides more meaningful and human-friendly translations which can be used in ad-hoc tasks apart from expert finding. For instance, recruiters can use these translations to select outstanding questions about a skill area.

\section{Related Work}
In this section, firstly, we review expert finding task and its methods in different domains. Then, we discuss prior works on CQA platforms. Our proposed models are inspired by semantic matching and translation models whose related studies also have been investigated in the last part of this section.

\subsection{Expert Finding}
In the past few years, expert finding has been attracted a lot of attention in the Information Retrieval community. As mentioned earlier, the task of expert finding is to retrieve and rank the experts given a field of expertise as an input query. 

This problem has been inquired in many environments such as organizations \cite{balog2009},  bibliographic networks \cite{neshati2014eg,Neshati:2012:MGF,DAUD2010615}, social networks \cite{Neshati_www,neshati2014eg}, Wikipedia \cite{ziaimatin2014expertise}, LinkedIn \cite{Budalakoti:2012}, CQAs \cite{neshati:2017:dynamicity,ROSTAMI2019} and even Instagram \cite{Pal:2016:insta}.

\par
There have been many studies on generative probabilistic models for this task which rank candidates according to $P(ca|q)$ indicating the probability of a candidate $ca$ being an expert in the topic $q$ (i.e. query) \cite{balog:trends}. These models are categorized into two groups including candidate generation models \cite{cao2005research,fang2007} and topic generation models \cite{balog2009}. Most of these methods mainly use raw textual pieces of evidence, ignoring domain-specific information (e.g. document quality or structure). Nevertheless, there are various methods proposed to extend and enhance expertise retrieval in many ways. Deng \textit{et al.} \cite{deng:enhanced} proposed a query-sensitive method to model the authors' authorities based on the community citation networks and developed an adaptive ranking method to enhance expertise retrieval. Furthermore, Zhao \textit{et al.} \cite{zhao2016expert} proposed a ranking metric network learning framework for expert finding using both users' relative quality rank to given questions and their social relations. These methods are somewhat failed to address the issue of vocabulary gap in expertise retrieval. However, there have been some studies to bridge this gap. Momtazi and Naumann \cite{Momtazi} proposed a topic modeling approach to extract the main topic of documents, then the extracted topics are acted as a bridge to find the probability of nominating each candidate as the expert for a given query. Additionally, Van Gysel \textit{et al.} \cite{VanGysel:2016} introduced an unsupervised discriminative model for this task by exclusively employing textual evidence via learning distributed word representations in an unsupervised way. 

\subsection{Community Question Answering}
Over the recent years, many studies have been done on detecting expert users in CQAs \cite{Pal:2015,Riahi:2012}. In these approaches, associated documents, social interactions, and the personal activities of each candidate are deemed as their expertise evidence. Nonetheless, CQA platforms are dynamic environments due to their immense daily posts, the rate of joining new users, changing in their activities and interests, emerging new topics and upward or downward trend (i.e. novelty) of topics.  Consequently, in these networks, experts should be detected not only by their textual pieces of evidence (i.e. documents), but also by using network structure and specific features of CQA \cite{pal2011early,derijke:2015:early,coldstart}.

Another aspect of research on CQAs is the automatic evaluation of the quality of user generated contents based on a defined measure (i.e. formula) \cite{blooma:2012,Ponzanelli:2014u}.

Neshati \textit{et al.} \cite{neshati:2017:dynamicity} also have introduced a new problem of detecting users that can be potential experts in future. The proposed method relies on the expertise evidence of users (i.e. documents) in current time and then according to these pieces of evidence, the authors have suggested a method to predict the best ranking of experts in future.

\subsection{Semantic Matching}
In recent years, a great deal of studies have been conducted in semantic matching as there are many approaches proposed in this task such as Query Reformulation \cite{Croft:2011:QRU}, Translation Models \cite{karimzadehgan:2010}, Topic Modeling \cite{blei2003latent} and etc. It should be noted that only Translation Models and Topic Modeling approaches are in the scope of this research study.

\par
Statistical machine translation (SMT) refers to statistical learning methods for translating texts from one language to another or the same language\cite{semantic}. To clarify, suppose ``CA'' as the main query. It is known that it can match ``California'' with a high degree of precision. In our problem, queries can be regarded as a single word (i.e. skill area), and documents are texts built from other words which have been used in that skill area. SMT technologies aim to deal with the mismatch between query and document in expert finding. As an instance, skill area ``java-ee" (i.e. query) can be translated to \textit{application, web, spring, bean, service, http, session, request, controller and ejb} which are crucial aspects of ``java-ee" in StackOverflow.

The primary idea of SMT methods is to estimate the probability of translating a document to a query. As a term can be translated to a set of other terms with a certain probability, SMT methods can address the vocabulary gap problem. Karimzadehgan and Zhai \cite{karimzadehgan:2010} have adopted a method to estimate statistical translation models (SMT) based on mutual information in which first off, the mutual information scores for each pair of words is calculated, and then the score is normalized to obtain a translation probability. 

As mentioned before, another method of semantic matching is topic modeling. Given a collection of documents, topic modeling techniques aim to discover the topics in the collection as well as the topic representations of the documents \cite{semantic}. One of the most popular methods for this approach is Latent Dirichlet Allocation (LDA) introduced by Blei \textit{et al.} \cite{blei2003latent}. It is by far the most widely used method in many machine learning, natural language processing, and information retrieval applications \cite{Momtazi}. Wei and Croft \cite{Wei:2006:LDM} applied this model to language model based information retrieval and compared it with probabilistic latent semantic indexing and cluster-based retrieval. Momtazi and Naumann \cite{Momtazi} have adopted this method for expert finding.

Over the past few years, there has been a growing trend towards Word Embedding (WE) approaches \cite{zou2013bilingual}. These models learn continuous-valued distributed representations of words known as embeddings \cite{mnih2013learning,pennington2014glove} in order to reduce the high dimensionality of words representations in contexts and increase generalization by introducing the expectation that similar word vectors signify semantically or syntactically similar words. WE has been used in many domains such as expert finding, \cite{VanGysel:2016}, product search \cite{VanGysel:2016:LLV} and etc.
In \cite{VanGysel:2016}, the authors have proposed an unsupervised semantic matching method for expert finding. They directly utilize the words as the features for expert finding task while in our WE approach we consider each skill area as a query and words are acting like a bridge between skill areas and candidates. Besides, we observed that query words are not commonly used in the body of posts. Therefore, we first find important words in each skill area (translation step), then we exploit this words to score candidates. Whereas, in \cite{VanGysel:2016} words are directly used for expert finding which would not be able to alleviate the vocabulary gap problem. Finally, In \cite{VanGysel:2016} the words are presented in one-hot representation which makes it challenging to run in domains with extensive vocabulary size, such as StackOverflow. To overcome this problem, we have utilized the LDA algorithm for word representation.

\section{Conclusion and Future work}
In this paper, we studied the problem of vocabulary gap between the expert finding query and terms which candidates use in their documents in StackOverflow. We first illustrated that by utilizing appropriate translations, we can overcome the mentioned gap. Additionally, a concept was defined in this study. Voteshare which exploits the way of identifying high-quality answers in the community. We used this concept to improve the precision of expert finding task. Then we proposed two translation models based on statistical co-occurrence of words and the word embedding approach. The main finding in this paper is that utilizing both the translation models and considering the quality of documents simultaneously can significantly improve the quality of expert finding in StackOverflow. Future work may target the diversification aspect of translation to select a diverse set of words for each query. Additionally, the effectiveness of the proposed models in other domains rather than programming CQAs could be analyzed in future works. Furthermore, models capable of translating each skill area into phrasal or multi-term words, are left for future work.




%\section*{References}
\bibliographystyle{elsarticle-num}
\bibliography{references}

%\begin{thebibliography}{00}
%%% Text of bibliographic item
%
%\bibitem{}
%
%\end{thebibliography}
\end{document}